\documentclass[aps,pra,twocolumn,reprint,showpacs,superscriptaddress,footinbib,groupedaddress]{revtex4}
\usepackage{dsfont}
\usepackage{amsmath}
\usepackage{graphicx}
\usepackage{multirow}
\usepackage[latin1]{inputenc}
\usepackage{ulem} 
\usepackage{units} 
\usepackage{version} 
\usepackage{color}
\usepackage{colortbl}
\usepackage{xcolor}  
\usepackage{fancyref}
\usepackage[colorlinks=true,citecolor=black,urlcolor=blue,linkcolor=black]{hyperref} 

\newcommand\stxt[1]{_{\text{#1}}} 
\newcommand\STXT[1]{^{\text{#1}}} 
\newcommand{\ms}{\ \text{ms}}

\newcommand{\Hz}{\ \text{Hz}}
\newcommand{\kHz}{\ \text{kHz}}

\newcommand{\mrad}{\ \text{mrad}}
\newcommand{\s}{\ \text{s}}

\newcommand{\beq}{\begin{equation}}
\newcommand{\eeq}{\end{equation}}

\begin{document}


\title{Stability enhancement by joint phase measurements in a single cold atomic fountain}

\author{M. Meunier}
 
\author{I. Dutta}
\author{R. Geiger}
\email{remi.geiger@obspm.fr}
\author{C. Guerlin}
\altaffiliation{Present affiliation: Laboratoire Kastler-Brossel, 24, rue Lhomond, 75231 Paris Cedex 05, France}
\author{C.L. Garrido Alzar}
\author{A. Landragin}
\email{arnaud.landragin@obspm.fr}
\affiliation{LNE-SYRTE, Syst\`emes de R\'ef\'erences Temps-Espace, Observatoire de Paris, CNRS, UPMC, 61 Avenue de l'Observatoire, 75014 Paris, France}

\date{\today}

\begin{abstract}

We propose a method of joint interrogation in a single atom interferometer which overcomes the dead time between consecutive measurements in standard cold atomic fountains. The joint operation enables for a faster averaging of the Dick effect associated with the local oscillator  noise in clocks and with  vibration noise in cold atom inertial sensors. Such an operation allows achieving the lowest stability limit due to atom shot noise. We demonstrate a multiple joint operation in which up to five clouds of atoms are interrogated simultaneously in a single setup.
The essential feature of multiple joint operation, demonstrated here for a micro-wave Ramsey interrogation, can be  generalized to go beyond the current stability  limit associated with  dead times in present-day cold atom interferometer inertial sensors. 

\end{abstract}



\maketitle

\section{Introduction}
\label{introduction}

Over the past two decades, important progress in cold atom physics has established atom interferometry (AI) as a unique tool for precision measurements of time and frequency, and of gravito-inertial effects. Atom interferometry is now addressing various applications ranging from precision measurements of fundamental constants \cite{Guena2012,bouchendira_state_2013,Rosi2014}, to inertial navigation \cite{Canuel2006,Geiger2011,Barrett2013}, to geophysics \cite{Jiang2012,Gillot2014} and has been proposed for gravitational wave detection (see, e.g. \cite{Dimopoulos2008PRD}). In order to address these promising applications beyond the strict scope of atomic physics, new methods must be formulated and demonstrated experimentally to use the full potentialities of AI. 
The main limitation of current cold atom interferometers is dead times between successive measurements, corresponding to the preparation of the atom source and the detection of the atoms at the output of the interferometer.

In cold atom or ion clocks, dead times lead to the well-known Dick effect where aliasing of the local oscillator noise results in a degradation of the clock short term sensitivity \cite{dick_local_1990}. 
Several experiments have previously demonstrated a way to bypass this effect in relative comparisons between two clocks \cite{Bize2000,Takamoto2011,Chou2011}, and in realizing a clock in the specific case of a continuous cold beam atomic source \cite{devenoges_improvement_2012}.
Zero-dead-time operation of two interleaved atomic clocks was recently demonstrated, resulting in a reduction of the contribution of the local oscillator noise \cite{biedermann_zero-dead-time_2013}. 
However, besides the relevance of this proof of principle experiment, this method used two different atomic clocks (interrogated by the same local oscillator) and therefore requires more experimental maintenance as well as control over more systematic effects. Moreover, considering applications to inertial sensors, it is required to interrogate successive atom clouds at the same location in order to reject all parasitic inertial terms, such as centrifugal accelerations of gradients of accelerations.


Here, we propose and demonstrate a new  method of joint interrogation of cold atom clouds in a single  atomic fountain which overcomes the dead time limitation in  atom interferometers of high sensitivity.
Our joint interrogation method is inspired from atom juggling methods originally introduced in the context of cold atom collisions in atomic fountain clocks \cite{legere_quantum_1998} and only realized so far for concurrent measurements \cite{hu_simultaneous_2011,Canuel2006,Rosi2014}.
With an innovative and simple control sequence, we demonstrate the joint interrogation of up to 5 cold atom clouds simultaneously, resulting in a long Ramsey interrogation time (800 ms), high sampling rate (up to 5 Hz) and leading to a faster reduction of the Dick effect.
As cold atom inertial sensors use more than two light pulses, rejection of the Dick effect associated with vibration noise in these sensors requires more than two clouds being interrogated simultaneously in the same setup. Our multiple joint operation proposes for the first time this possibility and demonstrates it experimentally in a multi-clock configuration.


\section{Principle of the joint operation and experiments}
Most cold atom interferometers such as clocks, accelerometers or gyroscopes are sequentially operated in a sequence of total duration $T_c$, and typically consisting of 3 main steps: \textit{(i)} atom trapping, cooling, and preparation; \textit{(ii)} N-microwave-or-light-pulse AI sequence (Ramsey-like interrogation with a total duration $T$); \textit{(iii)} atomic state detection. 
Here we present  experiments operating in joint mode ($T_c = T$) or multiple joint mode ($T/T_c=2,\ 3,\ 4$), resulting in a null dead time and enhanced stability of the interferometer.

 \begin{figure}[!h]
\includegraphics[width=\linewidth]{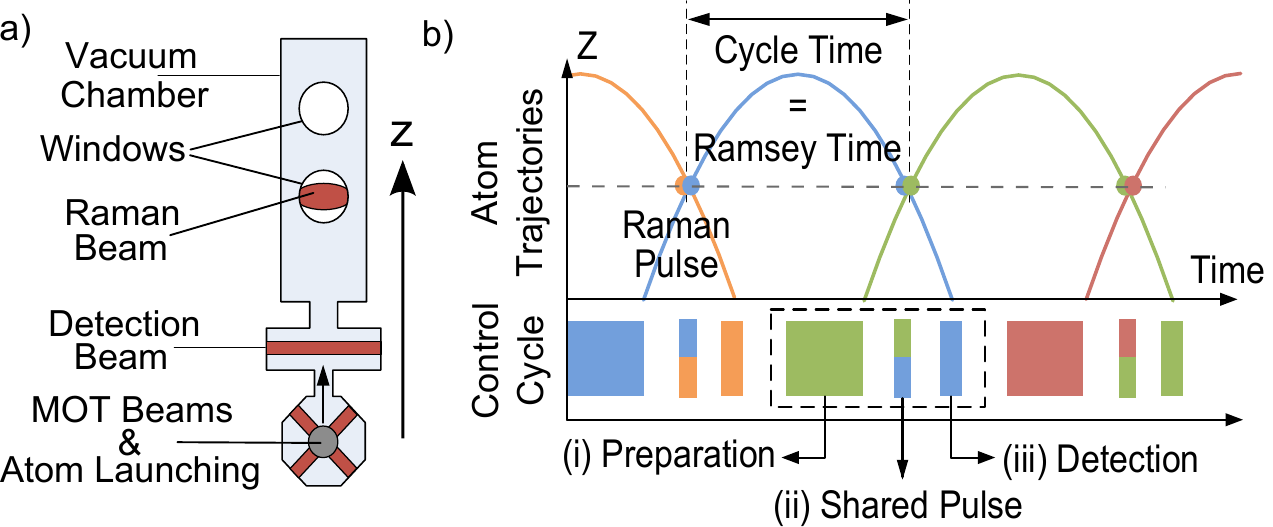}
\caption{\label{fig:jointM} 
(Color online)
a) Schematic of the instrument. 
b) Principle of the joint mode operation: \textit{(i)} preparation of cloud $N$, \textit{(ii)} Raman light pulse shared by clouds $N-1$ (falling) and $N$ (rising), and \textit{(iii) } detection of cloud $N-1$.}
\end{figure}

\label{par:exp}
The normal-mode interferometer is operated sequentially following the steps \textit{(i)-(iii)}, with an interrogation time $T=480 \ms$ and a cycle time $T_c=900 \ms$. Fig.\ref{fig:jointM}b) presents the principle of the joint mode operation where the Raman interrogation pulse is shared by clouds $N-1$ (falling) and $N$.
Fig.~\ref{fig:jointM}a) shows a schematic view of the experiment. Cesium atoms loaded from a 2D Magneto-Optical Trap (MOT) are trapped and cooled in a 3D-MOT; $4\times 10^7$ atoms  are launched vertically towards the interferometer region using moving molasses  with a temperature of  $1.3 \ \mu\text{K}$. The launching is followed by a microwave pulse selecting $6\times 10^6 $  atoms in the $|F=3,m_F=0\rangle$ state, which are used for interferometry. 
Light pulse interferometry is realized using co-propagating Raman lasers  which couple the  $|F=3,m_F=0\rangle$ and $|F=4,m_F=0\rangle$ clock  levels characterized by an hyperfine splitting of  $9.192 \ \text{GHz}$.
Several windows enable versatile configurations for the interferometer where interrogation times up to $800 \ms$ can be reached. In this work, we perform a Ramsey interrogation using two $\pi/2$ Raman pulses symmetric with respect to the apogee of the atom trajectory (see Fig.~\ref{fig:jointM}b)).
Experimentally, the joint operation implies to trap a cloud of atoms in the bottom part of the chamber, while another atom cloud is in the interferometer or detection regions. This is a form of juggling with the clouds  without re-capture \cite{legere_quantum_1998}. 

A microwave pulse prepares  atoms in the non-magnetic ($m_F=0$) state before the interferometer in order to maximise the interferometer contrast. The selection is performed on the $|F=4,m_F=+1\rangle \rightarrow |F=3,m_F=0\rangle$ transition which is separated from the clock transition using a bias field of  $18 \ \text{mG}$. This scheme allows avoiding perturbation of the atoms being interrogated in the Ramsey zone  by the microwave selection radiation.

\begin{figure}[!h]
\includegraphics[width=\linewidth]{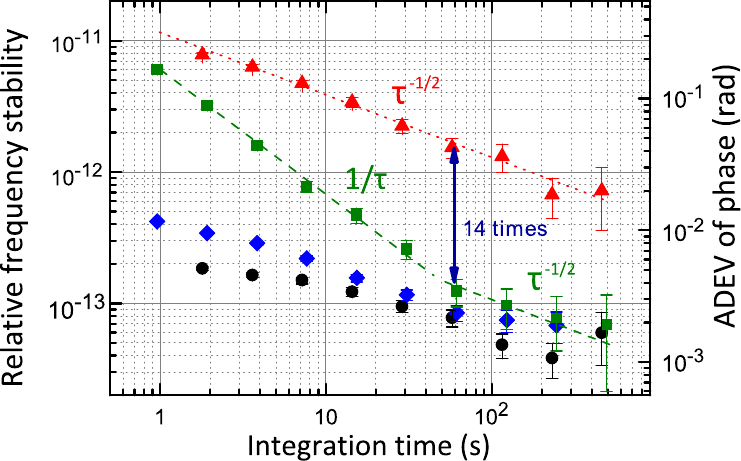}
\caption{\label{fig:LoFNoise}
(Color online) Allan deviations (ADEV) of the fountain relative frequency stability in normal and joint modes, for an interrogation time $T=480 \ms$. 
Stability without adding noise for the normal  (black circle) and joint  (blue rhombus) operations. 
 Allan deviation  for the normal mode (red triangle) and the joint mode (green square) when adding white noise over a bandwidth of $400 \Hz$. 
The $1/\tau$ (dashed) and $\tau^{-1/2}$ (dotted)  lines are guide to the eyes. 
We observe a 14-fold gain in frequency stability from normal to joint mode at $60 \ \text{s}$. 
Integrating to this frequency stability level in normal operation would require $12 \ 000 \ \text{s}$. 
}
\end{figure}

With $\pi/2$ pulses of duration $\tau_p=22 \ \mu \text{s}$ (Rabi frequency $\Omega_R/2\pi=11.4 \kHz$), the phase sensitivity extrapolated at $1 \ \text{s}$ of the 2-pulse Raman interferometer for the normal ($T=480 \ms, \ T_c=900 \ms$) and joint mode ($T_c=T=480 \ms$) operations   are  $13 \mrad$ and $16 \mrad$, respectively (see Fig.~\ref{fig:LoFNoise}, black circles and blue rhombus). The sensitivity is  limited by both the performance of the Raman laser phase lock system and by the detection noise.
The small difference in detection noise can be explained by fringe contrast loss from $50\%$ to $30\%$ when implementing the joint operation. This contrast loss  originates from the stray light scattered from the MOT atoms which interacts with the atoms in the interferometer region, starting $0.5$ m above, transferring them in unwanted states. Moreover, this stray light induces light shift on the interference fringes (see Appendix A, Fig.~\ref{fig:MultiFringes}). The effects of the MOT scattered light could be suppressed with the use of a vacuum compatible controllable shutter \cite{Fuezesi2007} between the MOT and interrogation regions.

\begin{figure*}[!ht]
\includegraphics[width=\linewidth]{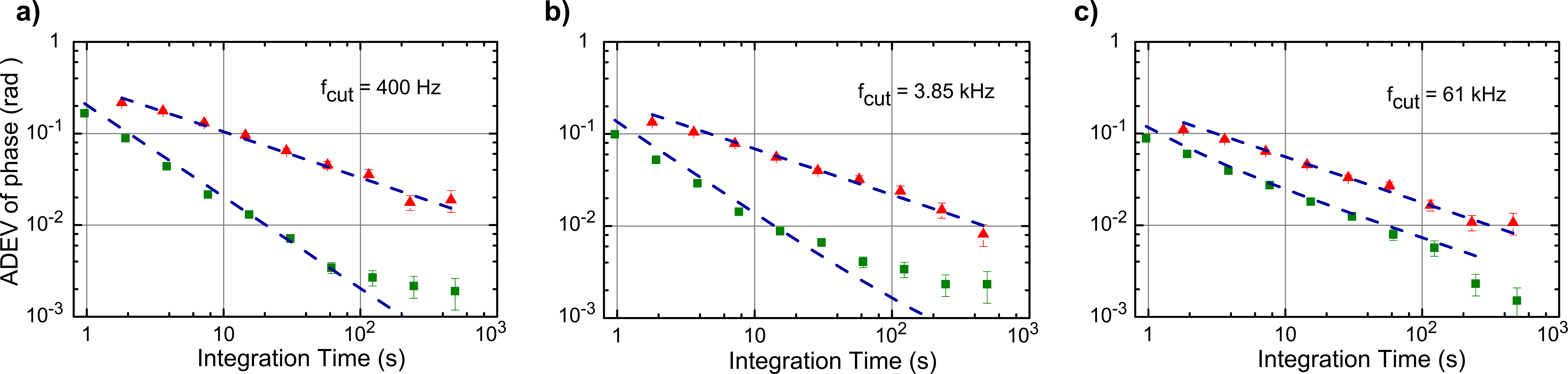}
\caption{\label{fig:FreqStudy} 
(Color online)
Comparison of normal mode (red triangles) and joint mode (green squares) for several cut off frequencies $f\stxt{cut}$ of added white noise to the Raman laser phase lock loop: $400 \Hz$ (a), $3.85 \kHz$ (b) and $61 \kHz$ (c). The Raman pulse Rabi frequency is $f_R = 11.4 \kHz$. 
Dashed blue lines: theoretical calculation based on Eq.~\eqref{eq:AVAR} without free parameter.
}
\end{figure*}

\section{Rejection of the local oscillator noise in joint operation}

\label{par:principle}
The  phase of the 2-pulse atom interferometer is determined by the Raman laser phase difference imprinted on the atomic wave-function at the light pulses;  at time $t_i$ it reads $\Delta \Phi_i$  = $\phi(T+t_i) - \phi(t_i)$, where $\phi(t)$ is the Raman laser relative phase. In the case of a white relative phase noise and after $N$ cycles, the variance $\langle\Delta \Phi_N^2\rangle$ of the  accumulated atomic phase is inversely proportional to $N$. In the time domain, this means that the phase Allan deviation scales as $1/\sqrt{\tau}$ ($\tau$ is the integration time), which is the well-known result for successive uncorrelated measurements.
With the cycle time $T_c$ equal to the Ramsey time $T$, the second laser pulse $\phi (T+t_i )$ of cloud $i$ is the same as the first pulse $\phi (t_{i+1})$ of cloud $i+1$: $\phi(T+t_i)=\phi(t_{i+1})$. As a result, the consecutive phase terms in the accumulated atomic phase cancel each other, so that the variance of the accumulated phase $\langle\Delta \Phi_N^2\rangle$ scales as $1/N^2$ (Allan deviation of phase $\sim 1/\tau$). 
In other words the joint operation rejects the aliasing of the local oscillator noise (here the Raman laser relative phase noise) usually encountered  when performing independent measurements of the phase with dead times. 
The rejection applies as long as the local oscillator noise spectrum has a bandwidth lower than the pulse Rabi frequency $\Omega_R$. We  quantitatively analyze the rejection efficiency below.

To demonstrate the local oscillator (LO) phase noise rejection, we introduce a white noise  of controlled amplitude and bandwidth in the Raman laser phase lock loop. The noise is generated using a Direct Digital Synthesizer (SRS DS345) and filtered by an analog $115 \ \text{dB}/\text{octave}$ low-pass filter (SR 650). The spectrum of added noise and the details of its calibration are given in the  Appendix B, Fig.~\ref{fig:Added_Noise}. 
Fig.~\ref{fig:LoFNoise} shows the measured phase Allan deviation (ADEV) for a white noise of $400 \Hz$ bandwidth (green squares), well below the Rabi frequency of $11.4 \kHz$. We clearly observe the expected $1/\tau$ scaling of the joint operation. 
A change of slope in the ADEV is observed at $60 \s$ when reaching the uncorrelated noise floor  at a level of $1\times10^{-13}$ in relative frequency stability,  corresponding to detection noise. 
Fig.\ref{fig:LoFNoise} thus show that the joint operation allows fast averaging to the fundamental noise linked to the detection noise even with a low stability local oscillator.

\label{par:quantitative_analysis}
The joint operation efficiently rejects the Dick effect associated with low frequencies in the LO noise, but the rejection is less efficient for  LO noise  bandwidths $f\stxt{cut}$ higher than the Raman pulse Rabi frequency $f_R=\Omega_R/2\pi$. 
In the following we explore the limits of the rejection.
Fig.~\ref{fig:FreqStudy} presents phase ADEV for measurements corresponding to LO noise bandwidths of  $400 \Hz$ (a), $3.85 \kHz$ (b) and $61 \kHz$ (c). 
The $1/\tau$ region expands over longer interrogation times for $f\stxt{cut}=400 \Hz$   than for $f_c=3.85 \kHz$. In the latter case, the Allan deviation changes its slope after $\sim 10 \s$ of integration time. In the $61 \kHz$ case, the $1/\tau$ scaling is no longer visible:  the joint mode no longer samples the LO noise so that there exists no correlation between successive measurements. 

To quantitatively analyse our data, we use the AI sensitivity function formalism, which provides the response of the atom interferometer to a perturbation at a given frequency  \cite{Lemonde1998,cheinet_measurement_2008}. The Allan variance of the phase  reads:
\begin{equation}
\sigma ^2 (\tau)=\frac{1}{2m^2} \int _0 ^{+\infty} \frac{d\omega}{2\pi} \vert H(\omega) \vert ^2 S_{\phi} (\omega) \frac{4\sin ^4 (m\omega T_c/2)}{\sin ^2 (\omega T_c /2)}
\label{eq:AVAR}
\end{equation}
where $\vert H(\omega) \vert ^2$ is the interferometer sensitivity function, $S_{\phi} (\omega)$ is the  noise power spectral density of the Raman laser relative phase, and $\tau=m T_c$; $m$ is therefore the number of averaged samples in the calculation of the Allan variance.
The 2-pulse inteferometer sensitivity function is given by \cite{Lemonde1998,cheinet_measurement_2008}:
\beq
\vert H(\omega) \vert ^2 = \frac{4\omega^2\Omega_R^2}{(\omega^2-\Omega_R^2)^2}\Big [ \cos\omega(\frac{T}{2}+\tau_p) + \frac{\Omega_R}{\omega}\sin\frac{\omega T}{2}\Big]^2
\label{eq:sensFunc}
\eeq
with $\tau_p$ the duration of the Raman $\pi/2$ pulse. Using the measured white Raman phase noise levels $S_\phi(\omega)=S_0$ (Fig.~\ref{fig:Added_Noise}) and evaluating Eq.~\eqref{eq:AVAR} numerically, we obtained the dashed lines in Figs.~\ref{fig:FreqStudy} a)-c) for the normal mode ($T_c=900 \ms$, $T=480 \ms$) and for the joint mode ($T_c=T+\tau_p$). 
Our calculation reproduces well the experimental results, without free parameter. In particular, the change of slope from $\tau^{-1}$ to $\tau^{-1/2}$ is well captured. It occurs at the point in time when the contribution of the high frequency noise ($f\stxt{cut}>f_R$) starts overcoming the low frequency noise contributions ($f\stxt{cut}<f_R$) which are well correlated in successive joint measurements.

\begin{figure}[!h]
\includegraphics[width=0.9\linewidth]{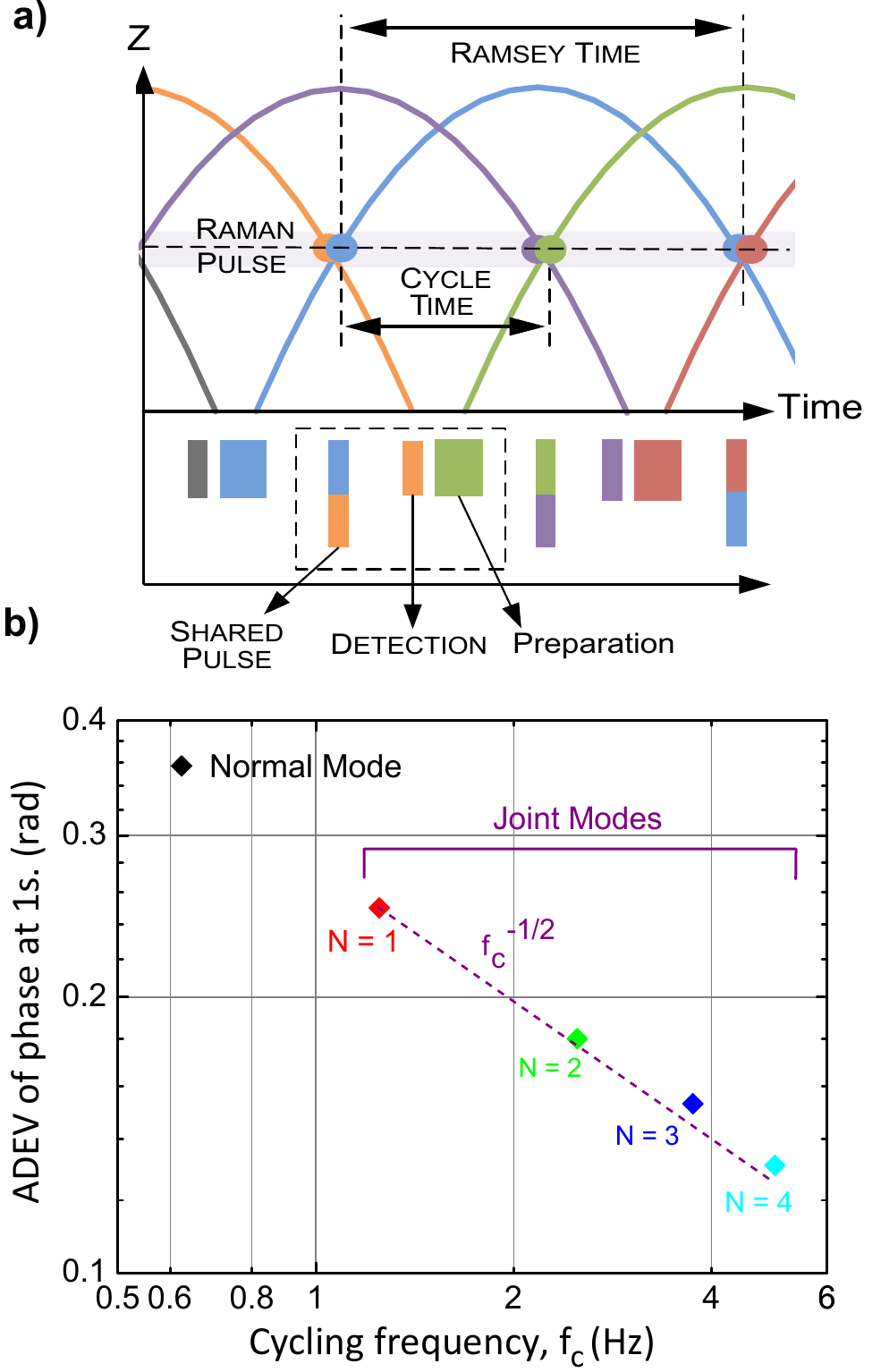}
\caption{
(Color online)
a) Schematic of the double joint operation where $T/T_c=2$,  where three atom clouds  simultaneously interact with the Raman laser pulses. 
b) Short term sensitivity at 1 second for each of the operation modes and $T=801 \ms$, from the normal operation ($T_c = 1.6 \ \text{s}$) to the quadrupole joint mode ($T/T_c=4$). The dashed line is a guide to the eye showing  the $1/\sqrt{f_c}$ scaling.}
\label{fig:MultipleJ}
\end{figure}

\section{Multiple Joint operation}

\label{par:multipleJoint}
We now present the extension of our method to a multiple joint operation where we interleave more than two atom interferometry measurements.
This ability is essential to reject the Dick effect associated with vibration noise in cold atom inertial sensors which use more than two light pulses to build the interferometer and thus require more than two clouds being jointly interrogated. 
The agility of our experimental setup allows us to further enhance the interferometric sensitivity  by juggling with more than two atom clouds, resulting in a cycle time $T_c$ being a sub-multiple of the Ramsey time $T$. Increasing $T$ to $801 \ms$, we present four  configurations of the joint operation with $T/T_c = 1 -4$.
Fig.~\ref{fig:MultipleJ}a) presents the principle of the double joint configuration where $T/T_c$ = 2. 
To characterize  the sensitivity gain of the multiple joint operation, we proceed as before by introducing a $400 \Hz$ bandwidth white noise to the Raman laser phase lock loop. 
For the four different configurations of $T_c$=[801, 400.5, 267, 200.25] ms, we observed similar $1/\tau$ scaling of the phase ADEV in the different joint modes. 
Fig.~\ref{fig:MultipleJ}b) shows the short term sensitivity at 1 s versus the cycling frequency $f_c=1/T_c$. 
For the single joint mode  interferometer with $T=801 \ms$, the short-term phase sensitivity is  $250 \mrad$, while it is $131 \mrad$ in  quadrupole  joint mode. This demonstrates a sensitivity enhancement of $1.9$ close to the expected value $(f_c\STXT{quad}/f_c\STXT{single})^{1/2}=2$. 

The minimum value of the cycle time in the multiple joint mode is  limited by the duration of preparation of the cold atoms. In our setup, we used a $150 \ms$ long MOT loading stage where the detection noise is at the limit between quantum projection noise and technical noise. Going beyond  could be achieved with optimum loading of the MOT.

\section{Conclusion}
\label{conclusion}
We have demonstrated a new method for operating an atom interferometer in a joint configuration where five atom clouds are interrogated simultaneously by common interrogation lasers. Our method highly rejects the Dick effect present in standard atomic fountains operated with dead times. It enables faster averaging of the phase noise  to achieve the fundamental noise floor linked to the detection noise.
This method remains efficient as soon as the noise frequency components are below the pulse Rabi frequency.  
 We also demonstrated an extension of this method to a multiple joint scheme, where several interleaved interrogations result in further improvement of the stability. 

The joint method enables to run microwave frequency standards at best performances (i.e. at the quantum projection noise limit) without sophisticated and expensive ultra-stable oscillator \cite{Grop2010,Hartnett2012,FemtoSTwebsite} by canceling the Dick effect (and increasing the locking bandwidth of the local oscillator).
Our method can be directly used for rejection of parasitic vibrations in cold atom inertial sensors operated with dead times. In such systems in which the signal is fluctuating, dead times not only reduce drastically the stability but lead to loss of information \cite{Jekeli2005}. Moreover, the multiple joint operation gives access to high frequency components while maintaining high sensitivity linked to long interaction times achievable with cold atom sensors.

\section{Acknowledgements.}
\label{par:acknowledgements}
We thank Bertrand Venon, Thomas L\'ev\`eque, David Horville and Michel Lours for their contributions to the design of the experiment   and the first stages of the realization, and Denis Savoie for its contribution to the simulations.
We thank DGA (D\'el\'egation G\'enerale pour l'Armement), IFRAF (Institut Francilien de Recherche sur les Atomes Froids) for funding. 
 M.M was supported by D\'el\'egation G\'en\'erale pour l'Armement (DGA), and I.D. by Centre National d'Etudes 
Spatiales (CNES) and LABEX Cluster of Excellence FIRST-TF (ANR-10-LABX-48-01) within the Program "Investissements d'Avenir" operated by the French National Research Agency (ANR).

M.M and I.D. contributed equally to this work.

\bibliography{Meunier_main_text}
\label{bibliography}


\section{Appendix}

\subsection{Influence of the light scattered by the atoms in the MOT}

\begin{figure}[!h]
\includegraphics[width=\linewidth]{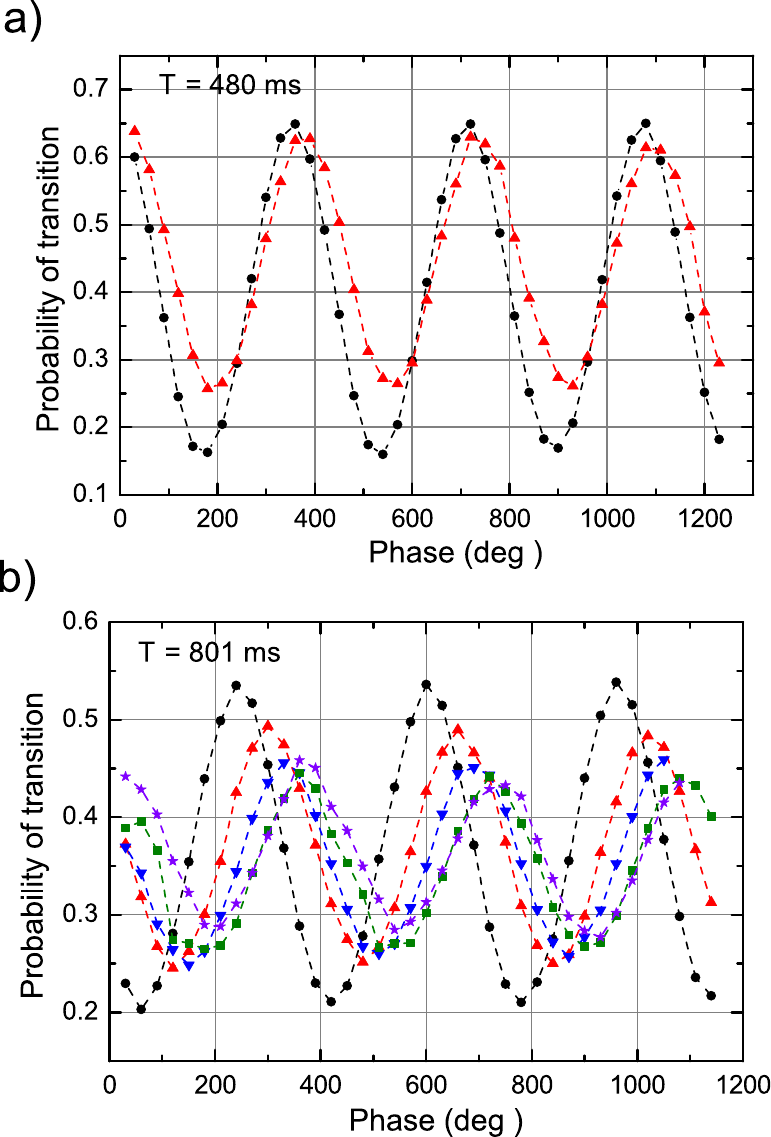}
\caption{
(Color online)
Interference fringes for two different Ramsey interrogation times: a) $T=480 \ms$ for the normal (black dots) and joint (red triangles) modes ; b) $T=801 \ms$ for the normal (black dots), joint (red triangles) and multiple joint operations (double joint: blue triangles, triple joint: green squares, quadrupole joint: violet stars).}
\label{fig:MultiFringes}
\end{figure}

See Fig.~\ref{fig:MultiFringes}.

\subsection{Added Raman laser phase noise}

\begin{figure}[!th]
\includegraphics[width=\linewidth]{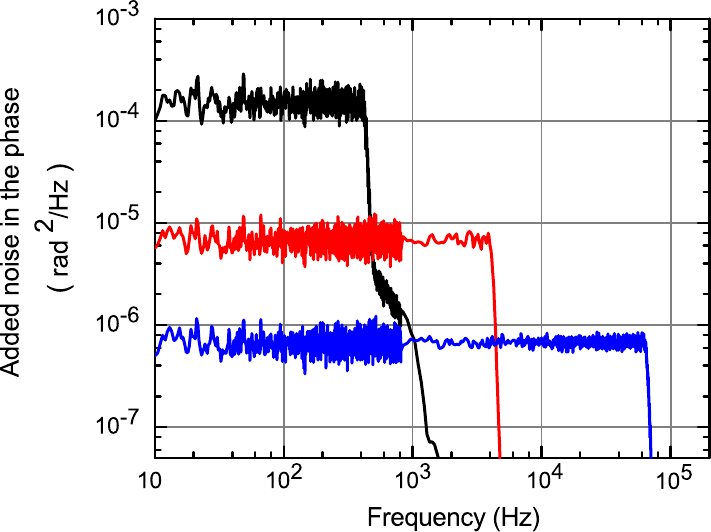}
\caption{ 
(Color online)
Power spectral density of the white added noise for the three different values of $f_{cut}$.
Black (top): $f_{cut}$ = 400 Hz; Red (middle): $f_{cut}$ = 3.85 kHz; Blue (bottom): $f_{cut}$ = 61 kHz. 
}
\label{fig:Added_Noise}
\end{figure}

The white noise  was generated using a Direct Digital Synthesizer (SRS DS345) and filtered by an analog $115 \ \text{dB}/\text{octave}$ low-pass filter (SR 650), and directly added to the Raman laser phase lock loop. The spectrum of the added noise is given in Fig.~\ref{fig:Added_Noise}. 
Its exact calibration  was important for the quantitative analyses presented in Fig. 3 of the main text.
To calibrate the noise level, we applied a sinusoidal modulation of known amplitude (in Volts) and frequency ($0.1 \Hz$) to the Raman laser phase lock loop and measured the  corresponding modulation of the interferometer phase (in radians). This yielded the conversion factor from Volts to radians and  the following white noise levels:
\newline
\newline
	$S_0 = 1.5\pm0.3\times 10^{-4} \ \text{rad}^2/\text{Hz}$ for $400 \Hz$ ; \newline
	$S_0 = 6.8\pm1.4\times 10^{-6} \ \text{rad}^2/\text{Hz}$ for $3.85 \kHz$ ;\newline
	$S_0 = 6.7\pm1.3\times 10^{-7} \ \text{rad}^2/\text{Hz}$ for $61 \kHz$.	
	\newline
	\newline
These measured noise levels were used in Eq.~\eqref{eq:AVAR}.
	
\end{document}